\documentclass[
    ,final            
  ]
  {aipproc}

\layoutstyle{8x11double}

\begin{document}

\title{On the Intermediate Subgroup of the Gamma-Ray Bursts in the Swift Database}

\classification{01.30.Cc, 95.55.Ka, 95.85.Pw, 98.70.Rz}
\keywords      {gamma-ray astrophysics, gamma-ray bursts}

\author{David Huja}{address={Astronomical Institute of the Charles University, V Hole\v{s}ovi\v{c}k\'{a}ch 2, Prague, Czech Republic}}

\author{Attila M\'esz\'aros}{address={Astronomical Institute of the Charles University, V Hole\v{s}ovi\v{c}k\'{a}ch 2, Prague, Czech Republic}}

\begin{abstract}

A sample of 286 gamma-ray bursts, detected by Swift satellite, is studied statistically by the $\chi^2$ test and the Student t-test, respectively. The short and long subgroups are well detected in the Swift data. But no intermediate subgroup is seen. The non-detection of this subgroup in the Swift database can be explained, once it is assumed that in the BATSE database the short and the intermediate subgroups form a common subclass.

\end{abstract}

\maketitle

\section{Data Samples}

According to our knowledge, gamma-ray bursts (GRBs) are the most powerful mysterious explosions the Universe has ever seen since the Big Bang. With Swift satellite, since November 20, 2004, we have a tool, which can solve the gamma-ray burst mystery. We define two samples from the Swift dataset \cite{1}: The sample of GRBs without measured redshift $z$ (189 GRBs), and the sample with measured $z$ (97 GRBs). The Swift catalogue consists of the name of GRB, its BAT duration $T_{90}$, BAT fluence at range 15-150 keV, BAT peak flux at range 15-150 keV, and redshift. The sample covers the period November 2004 - December 2007; the first / last event is GRB041227 / GRB071227. We have studied both samples separately and also together as the whole sample (286 GRBs).

\section{$\chi^2$ fitting of the duration for the whole sample}

The first evidence about the existence of three subgroups of GRBs came from the $\chi^2$ fitting of the durations of BATSE dataset \cite{3}. We proceed here identically. We study the whole sample and the two ones. On the x-axis there are the bins (intervals) of decimal log~$T_{90}$, and on the y-axis there is the number of GRBs in each interval. The number of bins is 10 and the bins define a histogram. We fitted the histogram. The whole sample of the GRBs has 276 GRBs with the measured duration. The best fit of the whole sample with one single Gaussian curve gives: $\mu$ = 1.42 (with mean $T_{90}$ = 26.30s ), $\sigma$ = 0.94, $\chi^2$ = 53.54. The goodness of the fit with 8 $dof$ (degrees of freedom) gives the rejection on the 99.99\% significance level \cite{4,5}. The fit with the sum of two Gaussian curves gives: $\mu_{1}$ = -0.27 ($T_{90}$(1) = 0.54s) ,  $\sigma_{1}$ = 0.94,  $\mu_{2}$ = 1.57 ($T_{90}$(2) = 37.15s), $\sigma_{2}$ = 0.56, $w$ = 0.12 (12\% of the GRBs belongs to the short GRBs), $\chi^2$ = 6.72. Here $\chi^2$ $\simeq$ $dof$ = 5 and we obtain an excellent fit with the significance level 50\%. The fit with the sum of three Gaussian curves gives: $\mu_{1}$ = 0.30 ($T_{90}$(1) = 2.00s),  $\sigma_{1}$ = 1.27, $w_{1}$ = 0.20, $\mu_{2}$ = 0.82 ($T_{90}$(2) = 6.61s),  $\sigma_{2}$ = 0.07, $w_{2}$ = 0.12, $\mu_{3}$ = 1.72 ($T_{90}$(3) = 52.48),  $\sigma_{3}$ = 0.46,  $\chi^2$ = 3.82, we obtain an excellent fit for $dof$ = 2, because the significance level is only 78\%. The decreasing of $\Delta \chi^2$ = 2.89 is not statistically significant \cite{4,5}, and hence the introduction of the third intermediate subgroup is not necessary. The fit is shown on the Figure 3.

\begin{figure}
  \includegraphics[width=0.7\textwidth]{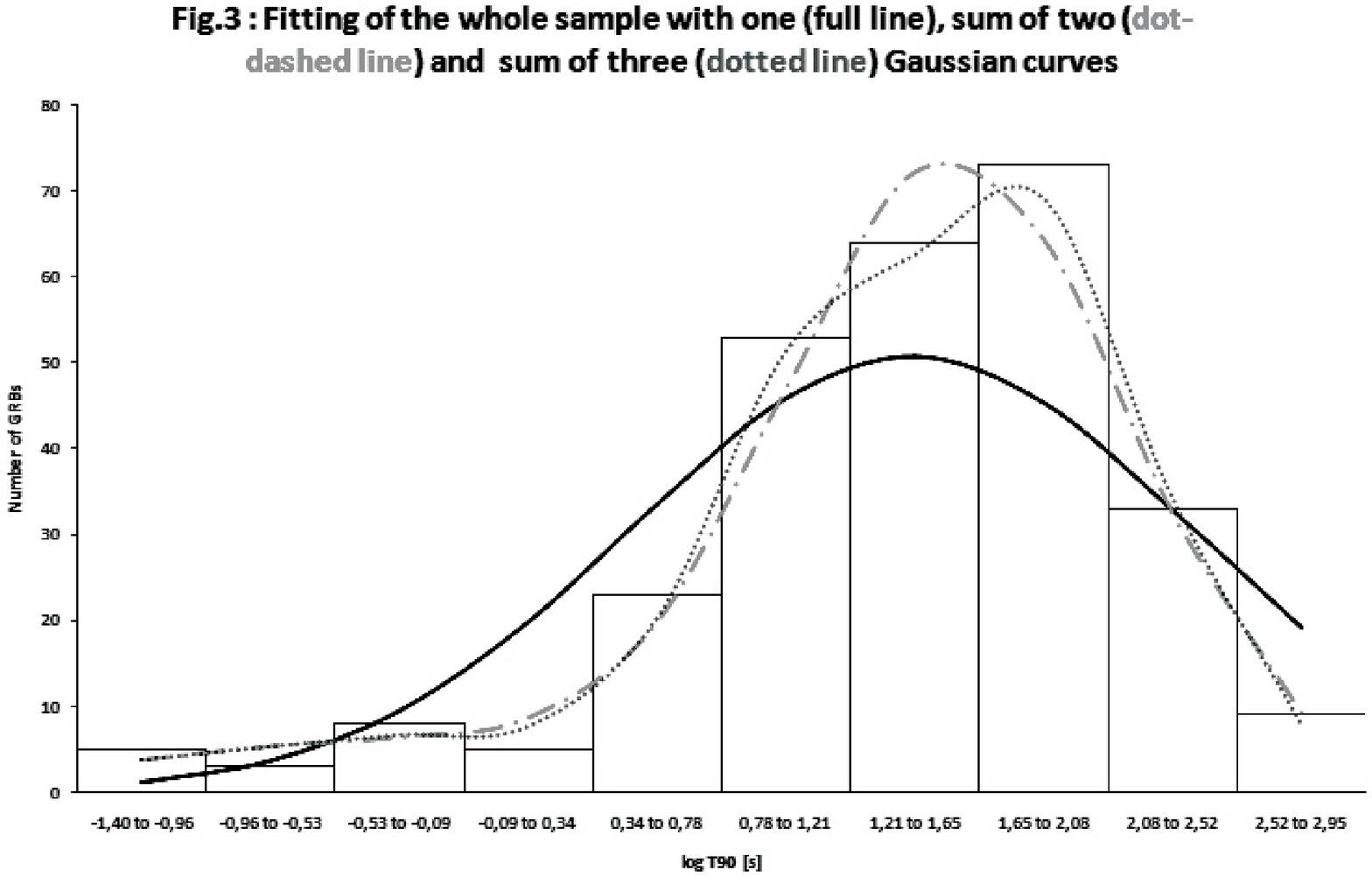}
\end{figure}

\section{$\chi^2$ fitting of the duration for the sample with measured redshift}

The sample with measured redshifts contains 94 GRBs with measured duration. The best fit of the sample with one single Gaussian curve gives: $\mu$ = 1.43 ($T_{90}$ = 26.92s), $\sigma$ = 0.87, $\chi^2$ = 12.59. The goodness of the fit with 8 $dof$ does not reject the hypothesis of one single Gaussian curve. The rejection is only on the 85\% significance level. The fit with the sum of two Gaussian curves gives: $\mu_{1}$ = -0.46 ($T_{90}$(1) = 0.35s), $\sigma_{1}$ = 0.51,  $\mu_{2}$ = 1.52 ($T_{90}$(2) = 33.11s), $\sigma_{2}$ = 0.62, $w$ = 0.06 (6\% of the GRBs belongs to the short GRBs), $\chi^2$ = 3.45. Here we obtain an excellent fit with the significance level 97\% \cite{4,5}, thus the fit can not be rejected. The fit with the sum of three Gaussian curves gives: $\mu_{1}$ = -0.71 ($T_{90}$(1) = 0.19s),  $\sigma_{1}$ = 0.70, $w_{1}$ = 0.09, $\mu_{2}$ = 0.51 ($\chi^2$(2) = 3.24s), $\sigma_{2}$ = 0.01, $w_{2}$ = 0.01, $\mu_{3}$ = 1.56 ($T_{90}$(3) = 36.31s), $\sigma_{3}$ = 0.54, $\chi^2$ = 3.16, it is an excellent fit with $dof$ = 2, because the significance level is only 80\%. The decreasing of $\Delta \chi^2$ = 0.29 is not statistically significant \cite{4,5}, also the introduction of the third intermediate subgroup is not necessary. The fit is shown on the Figure 2.

\begin{figure}
  \includegraphics[width=0.7\textwidth]{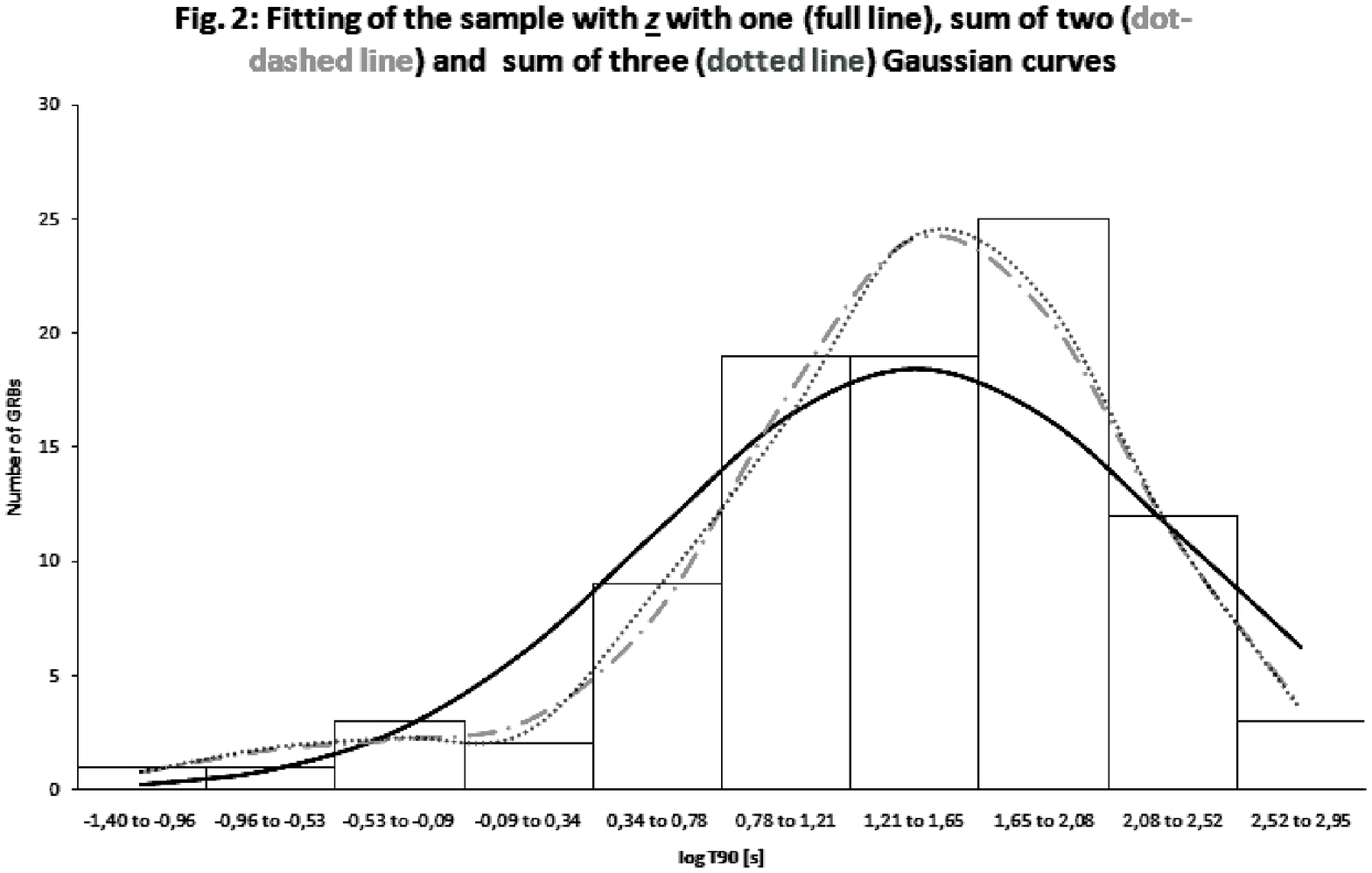}
\end{figure}

\section{$\chi^2$ fitting of the duration for the sample without measured redshift}

The sample without measured redshifts contains 182 GRBs with measured duration. The best fit of the sample with one single Gaussian curve gives: $\mu$ = 1.40 ($T_{90}$ = 25.12s), $\sigma$ = 0.95, $\chi^2$ = 43.23. The goodness of the fit with 8 $dof$ gives the rejection on the 99.99\% significance level. The fit with the sum of two Gaussian curves gives: $\mu_{1}$ = -0.73 ($T_{90}$(1) = 0.19s), $\sigma_{1}$ = 0.71, $\mu_{2}$ = 1.54 ($T_{90}$(2) = 34.67s), $\sigma_{2}$ = 0.56, $w$ = 0.09 (9\% of the GRBs belongs to the short GRBs), $\chi^2$ = 3.42. Here we obtain an excellent fit, which can not be rejected, \cite{4,5}. The fit with the sum of three Gaussian curves gives:  $\mu_{1}$ = -0.70 ($T_{90}$(1) = 0.20s),  $\sigma_{1}$ = 0.70, $w_{1}$ = 0.09,  $\mu_{2}$ = 0.51 ($T_{90}$(2) = 3.24s),  $\sigma_{2}$ = 0.01, $w_{2}$ = 0.01,  $\mu_{3}$ = 1.56 ($T_{90}$(3) = 36.31s),  $\sigma_{3}$ = 0.34, $\chi^2$ = 3.16, hence is an excellent fit with $dof$ = 2, because the significance level is only 80\%.The decreasing of $\Delta \chi^2$ = 0.26 is not statistically significant \cite{4,5}, and then the introduction of the third intermediate subgroup is not necessary. The fit is shown on the Figure 1.

\begin{figure}
  \includegraphics[width=0.7\textwidth]{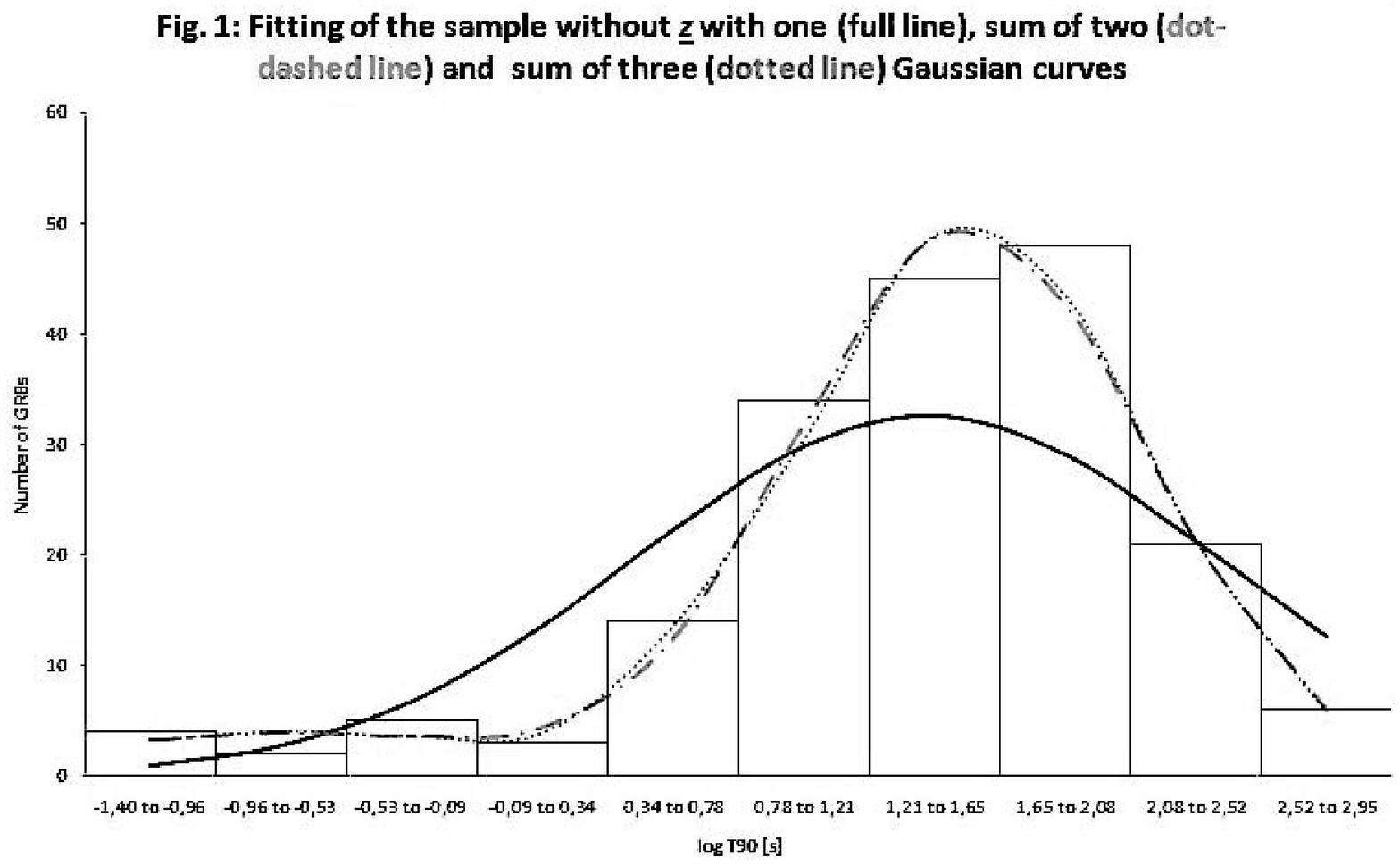}
\end{figure}

\section{Conclusion}

Because the article \cite{3} described the existence of the third (intermediate) subgroup of GRBs in the BATSE database by the $\chi^2$ fitting of the duration, we worked out an identical procedure on the existing Swift database. Contrary to the BATSE GRBs (\cite{3,6,7}), the Swift GRBs do not require any introduction of the third intermediate subgroup in contrast to \cite{8}; the hypothesis is that the short and intermediate subgroups should form one single subclass.

\begin{theacknowledgments}

This study was supported by the GAUK grant No.46307, by the OTKA grant No.T48870, by the Grant Agency of the Czech Republic, grants No. 205/08/H005 , and by the Research program MSM0021620860 of the Ministry of Education of the Czech Republic. The authors appreciate valuable discussion and help of J. \v{R}\'{\i}pa.

\end{theacknowledgments}

\end{document}